\documentclass[a4paper,12pt]{article}

\usepackage{refmerge}
\usepackage[dvips]{graphicx}
\usepackage{amssymb}

\newcommand{\etap}{\ensuremath{\eta^{\prime}}}
\newcommand{\etapg}{\ensuremath{\eta^{\prime}\gamma}}
\newcommand{\ks}{\ensuremath{K^0_S}}
\newcommand{\kl}{\ensuremath{K^0_L}}
\newcommand{\fourpi}{\ensuremath{\pi^+\pi^-\pi^+\pi^-}}

\title{Observation of the $\phi\to\etapg$ decay with four charged
  particles and photons in the final state}

\author{
R.R.~Akhmetshin\footnote{Budker Institute of Nuclear Physics,
  Novosibirsk, 630090, Russia},
E.V.Anashkin\footnotemark[1],
M.Arpagaus\footnotemark[1], \and
V.M.Aulchenko\footnotemark[1]
\footnote{Novosibirsk State University, Novosibirsk, 630090, Russia},
V.Sh.Banzarov\footnotemark[1], 
L.M.Barkov\footnotemark[1] \footnotemark[2], \and
N.S.Bashtovoy\footnotemark[1],
A.E.Bondar\footnotemark[1] \footnotemark[2],
D.V.Bondarev\footnotemark[1], \and
A.V.Bragin\footnotemark[1],  
D.V.Chernyak\footnotemark[1],
S.I.Eidelman\footnotemark[1] \footnotemark[2], \and
G.V.Fedotovitch\footnotemark[1] \footnotemark[2],  
N.I.Gabyshev\footnotemark[1], 
A.A.Grebeniuk\footnotemark[1], \and
D.N.Grigoriev\footnotemark[1],
V.W.Hughes\footnote{Yale University, New Haven, CT 06511, USA},
F.V.Ignatov\footnotemark[1] \footnotemark[2],
P.M.Ivanov\footnotemark[1], \and
S.V.Karpov\footnotemark[1],
V.F.Kazanin\footnotemark[1] \footnotemark[2],
B.I.Khazin\footnotemark[1] \footnotemark[2],
I.A.Koop\footnotemark[1], \and
M.S.Korostelev\footnotemark[1],
P.P.Krokovny\footnotemark[1] \footnotemark[2],
L.M.Kurdadze\footnotemark[1] \footnotemark[2], \and
A.S.Kuzmin\footnotemark[1] \footnotemark[2],  
I.B.Logashenko\footnotemark[1],
P.A.Lukin\footnotemark[1], \and
K.Yu.Mikhailov\footnotemark[1] \footnotemark[2],
I.N.Nesterenko\footnotemark[1], 
V.S.Okhapkin\footnotemark[1], \and
A.V.Otboev\footnotemark[1],
E.A.Perevedentsev\footnotemark[1] \footnotemark[2], 
A.S.Popov\footnotemark[1] \footnotemark[2], \and
T.A.Purlatz\footnotemark[1] \footnotemark[2], 
S.I.Redin\footnotemark[1],
N.I.Root\footnotemark[1] \footnotemark[2], \and
A.A.Ruban\footnotemark[1],
N.M.Ryskulov\footnotemark[1],
A.G.Shamov\footnotemark[1],  \and
Yu.M.Shatunov\footnotemark[1],
B.A.Shwartz\footnotemark[1] \footnotemark[2],
A.L.Sibidanov\footnotemark[1] \footnotemark[2], \and
V.A.Sidorov\footnotemark[1], 
A.N.Skrinsky\footnotemark[1], 
V.P.Smakhtin\footnotemark[1],\and
I.G.Snopkov\footnotemark[1], 
E.P.Solodov\footnotemark[1] \footnotemark[2],
P.Yu.Stepanov\footnotemark[1], \and
A.I.Sukhanov\footnotemark[1],
J.A.Thompson\footnote{University of Pittsburgh, Pittsburgh, PA 15260, USA},
V.M.Titov\footnotemark[1], \and
A.A.Valishev\footnotemark[1],
Yu.V.Yudin\footnotemark[1],
S.G.Zverev\footnotemark[1]
}

\date{\today}

\begin{document}

\maketitle

\newpage

\begin{abstract}
  Using 11.6 $\mbox{pb}^{-1}$ of data collected in the 
  energy range 0.864-1.06 GeV by CMD-2 at VEPP-2M, the rare decay mode
  $\phi\to\etapg$ was observed via the decay chain
  $\etap\to\pi^+\pi^-\eta$, $\eta\to\pi^+\pi^-\pi^0$ or
  $\eta\to\pi^+\pi^-\gamma$. The following branching ratio was obtained:
  \begin{displaymath}
    Br(\phi \to \etapg) = (4.9 ^{+2.2} _{-1.8} \pm 0.6) \cdot 10^{-5}
\,.
  \end{displaymath}
\end{abstract}

\section{Introduction}

Measurements of radiative decays of light vector mesons provide good
tests of the quark model and extend our understanding of  SU(3)
symmetry breaking \cite{odon,ben}. The discovery of the rare decay
$\phi\to\etapg$ by CMD-2 \cite{tanya1st} has brought the last piece
into the otherwise complete mosaic of radiative dipole magnetic
transitions between light vector and pseudoscalar mesons.
This observation was later confirmed by the SND group \cite{snd}.
Recently CMD-2 presented an improved measurement of the rate of
the $\phi\to\etapg$ decay \cite{tanya2nd} based on the total data
sample accumulated in the $\phi$-meson energy range.
The first preliminary results by the KLOE collaboration on the 
$Br(\phi \to \etapg)$ value reported recently \cite{kloe}
are consistent with those from CMD-2  and SND  
within the experimental uncertainties.

Because of the rather low probability of the studied decay as well as
serious background problems resulting in strict selection criteria,
none of  the above measurements have good statistical precision. One
way to increase a sample of $\eta' \gamma$ events is 
to  use other decay modes of the $\eta'$ and
$\eta$ mesons.      
In most of the previous papers \cite{tanya1st,snd,tanya2nd,kloe} events 
of the decay $\phi\to\etapg$ were selected in the mode 
$\etap\to\pi^+\pi^-\eta$, $\eta\to\gamma\gamma$. The pure neutral decay 
chain $\etap \to \pi^0\pi^0 \eta$, $\eta \to \gamma \gamma$ has been also 
used by the KLOE group \cite{kloe}.
In this paper we study the decay $\phi\to\etapg$ with the CMD-2 detector
at VEPP-2M in the channels:
\begin{eqnarray}
\phi \to \etapg\,,\quad
\etap \to \eta\pi^+\pi^-\,, &&
\eta \to \pi^+\pi^-\pi^0\quad \mbox{or}
\label{eq:3pi} \\
\phi \to \etapg\,,\quad
\etap \to \eta\pi^+\pi^-\,, &&
\eta \to \pi^+\pi^-\gamma. \label{eq:ppg}
\end{eqnarray}

The analysis is based on the data sample collected in three scans of 
the c.m.energy range from 0.984 to 1.060 GeV performed in winter 1997--1998.
The total integrated luminosity of
about 11.6 $\mbox{pb}^{-1}$ corresponds to approximately
16 million $\phi$ meson decays.

The general purpose detector CMD-2 has been described in detail 
elsewhere \cite{cmddec}. It consists of a drift chamber (DC) 
\cite{dcdec} and a 
proportional Z-chamber \cite{zcdec}, both used for the trigger, 
and both inside 
a thin (0.4 $X_0$) superconducting solenoid with a field of 1 T.

The barrel calorimeter \cite{csidec} which is placed outside the solenoid,
consists of 892 CsI crystals of $6\times 6\times 15$ cm$^3$ size and covers
polar angles from $46^\circ$ to $134^\circ$. The energy resolution for
photons is about 9\% in the energy range from 50 to 600 MeV. The angular
resolution is about 0.02 radians.

The end-cap calorimeter \cite{bgodec} which is placed inside the
solenoid, consists of 680 BGO crystals of $2.5\times 2.5\times 15$
cm$^3$ size and covers forward-backward polar angles from 16$^\circ$
to 49$^\circ$ and from 131$^\circ$ to 164$^\circ$. The energy and
angular resolution varies from 8\% to 4\% and from 0.03 to 0.02
radians respectively for the photons in the energy range from 100 to
700 MeV.

The luminosity was determined from Bhabha scattering
events at large angles \cite{prep99}.

\section{Data analysis}

The specific signature of the  processes
(\ref{eq:3pi}) and (\ref{eq:ppg}) is  the final state
with four charged pions coming from the interaction region and two
or more photons. It is well known that the $\phi$ meson decays 
do not provide such a signature with the exception of the following
decays into kaons:
\begin{eqnarray}
\phi &\to& \ks \kl ,\; \ks\to\pi^+\pi^- ,\;
\kl\to\pi^+\pi^-\pi^0 \quad\mbox{and}\label{eq:kskl} \\
\phi &\to& K^+ K^- ,\; K^{\pm} \to \pi^{\pm} \pi^+ \pi^- ,\;
K^{\mp} \to \pi^{\mp} \pi^0 .\label{eq:kpkm}
\end{eqnarray}
However, because of the large lifetime of kaons charged particles
originating from the kaon decays have a broad distribution of the
impact parameter with respect to the interaction point. It is this
feature which will be used to suppress  this background and
control the contamination  of the final event sample.
   
Another background reaction is $e^+e^- \to \omega\pi^0$, $\omega \to
\pi^+\pi^-\pi^0$ followed by the Dalitz decay of one of the pions:
$\pi^0 \to e^+e^-\gamma$. In this case the distribution
of the minimum space angle between the charged tracks has a
characteristic peak at small angles.

We selected events with four charged tracks coming from the
interaction region: the maximum impact parameter of the tracks
$ d = \max^{4}_{i=1} (r_{min_i})$ is less than 1 cm and
the vertex
coordinate along the beam axis $z_{vert}$ is within $\pm 10$ cm.
For better reconstruction efficiency, tracks were also required to
cross at least two superlayers of the drift chamber: $|\cos \theta| <
0.8 $. Two or more photons have to be detected in the calorimeter.
For the selected events the kinematic fit was performed taking into
account energy --- momentum conservation and assuming 
all charged particles to be pions.

One of the main problems in the analysis is additional (``fake'')
photons induced by the products of nuclear interactions of charged
pions in the detector material.
The following  method was used to suppress such fake photons.
In the kinematic fit the energy resolution of  
photons was loosened to the value $\sigma_{E_{\gamma}} = E_{\gamma}$ + 20 MeV. 
Thus, the photon energy was allowed to vary in a wide range during the fit. 
As the first step, the fit was performed in the hypothesis of the
$\fourpi\gamma\gamma$ final state.
For events with more than two detected photons ($N^{det}_{\gamma} > 2$)
the fit was repeated for all possible pairs of photons
and the pair with the smallest $\chi^{2}_{4\pi2\gamma}$ characterizing
the fit quality was selected.
Then, for these events, the assumption of the
$\fourpi\gamma\gamma\gamma$ state was tested.
For events with $N^{det}_{\gamma} > 3$, the combination of three
photons with the smallest $\chi^{2}_{4\pi3\gamma}$ was selected.
Finally, the signature $\fourpi\gamma\gamma$ or
$\fourpi\gamma\gamma\gamma$ was assigned to the event depending on whether
the magnitude of $\chi^{2}_{4\pi2\gamma}$ or
$\chi^{2}_{4\pi3\gamma}$ was smaller.
Then the requirement on the fit quality $\chi^2 = \min
(\chi^{2}_{4\pi2\gamma}, \chi^{2}_{4\pi3\gamma})$ was applied:
$\chi^2/\mbox{n.d.f} < 10/4$.
Events with the reconstructed photon energy below 30 MeV were rejected
from the subsequent analysis.

The background coming from the decay $\phi \to K^+ K^-$, in which
products of kaon nuclear interaction scatter back to the drift chamber
and induce two extra tracks or one of the kaon decays via the $K^{\pm}
\to \pi^{\pm} \pi^+\pi^-$ channel, accompanied by fake photons,
was suppressed using ionization losses of the tracks (see
\cite{dcdec,phi4pi} for more detail). Requiring $E_{4\pi}/2 E_{beam} >
0.7$, where $E_{4\pi} = \sum_{i=1}^{4}
\sqrt{\vec{p}^{2}_{i} + m^{2}_{\pi}}$ is the total energy of the
tracks, additional suppression of this type of the background was
achieved.

\begin{figure}
\begin{center}
\includegraphics[width=0.6\textwidth]{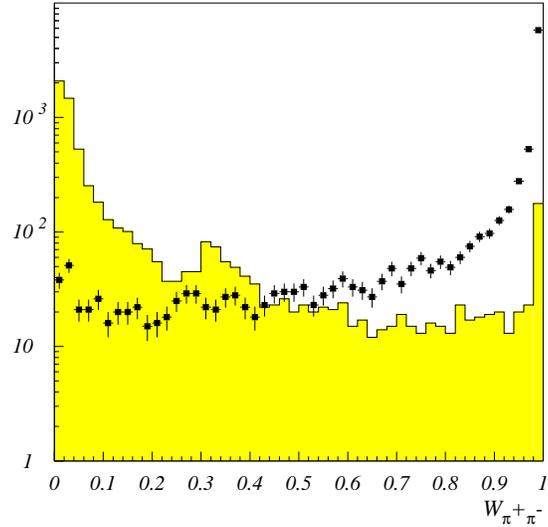}
\caption{$e/\pi$ separation probability. Distribution of $W_{\pi^+\pi^-}$
for the events of the processes $e^+e^- \to e^+e^-\gamma$ (histogram)
and $e^+e^- \to \pi^+\pi^-\pi^0$ (points with errors)}
\label{fig:epsep-phi}
\end{center}
\end{figure}
 
To suppress the background reaction $e^+e^- \to \omega \pi^0$, where
$\omega \to \pi^+ \pi^- \pi^0$ and one of the neutral pions decays to
$e^+e^-\gamma$, a previously tested procedure for $e/\pi$-separation 
(see \cite{rho4pi,conv} for details) was employed. In this
procedure, using the data
samples of $e^+e^- \to \pi^+\pi^-\pi^0$ and $e^+e^- \to
e^+e^-\gamma$ events, the probability density functions for the
parameter $E_{CsI}/|\vec{p}|$ were obtained for both particle types
$(e/\pi)$ and signs $(+/-)$: $f_{\pi^+}$, $f_{\pi^-}$, $f_{e^+}$ and
$f_{e^-}$. Here $E_{CsI}$ is the energy deposition in the CsI
calorimeter and $\vec{p}$ is the momentum of the particle with a track
matching the cluster in CsI. Then the probability for the pair of
tracks to be pions was defined: $W_{\pi^+\pi^-} = f_{\pi^+} f_{\pi^-}
/ ( f_{\pi^+} f_{\pi^-} + f_{e^+} f_{e^-})$. Figure \ref{fig:epsep-phi}
shows the distribution of $W_{\pi^+\pi^-}$ for the events of the processes
$e^+e^- \to e^+e^-\gamma$ and $e^+e^- \to \pi^+\pi^-\pi^0$ demonstrating
clear separation of the two types of the processes. 
Since the decay $\pi^0 \to e^+e^-\gamma$ features a small space
angle between $e^+e^-$, we searched for a pair of oppositely charged
particles with a minimum space angle $\psi_{min}$. The probability
$W_{\pi^+\pi^-}$ was calculated for this pair.
After that, the condition $W_{\pi^+\pi^-} > 0.5$ suppressed about 70\%
of the $\omega\pi^0$ events.
In addition, ionization losses $dE/dx$ were used to reject $\omega\pi^0$ 
events with a low momentum of $e^+e^-$ \cite{phi4pi}.

\begin{figure}
\begin{center}
  \includegraphics[width=0.49\textwidth]{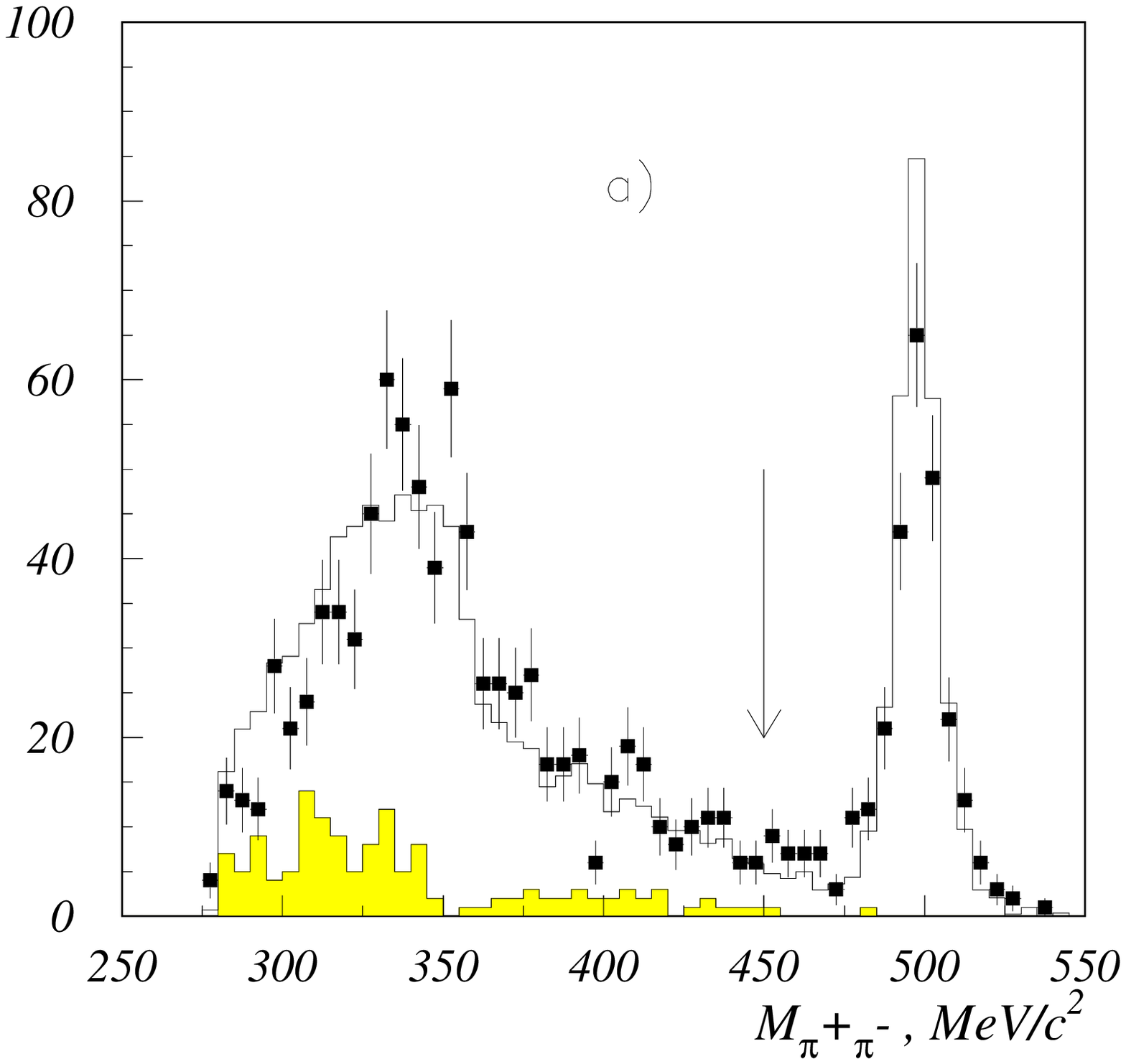}
  \hfill
  \includegraphics[width=0.49\textwidth]{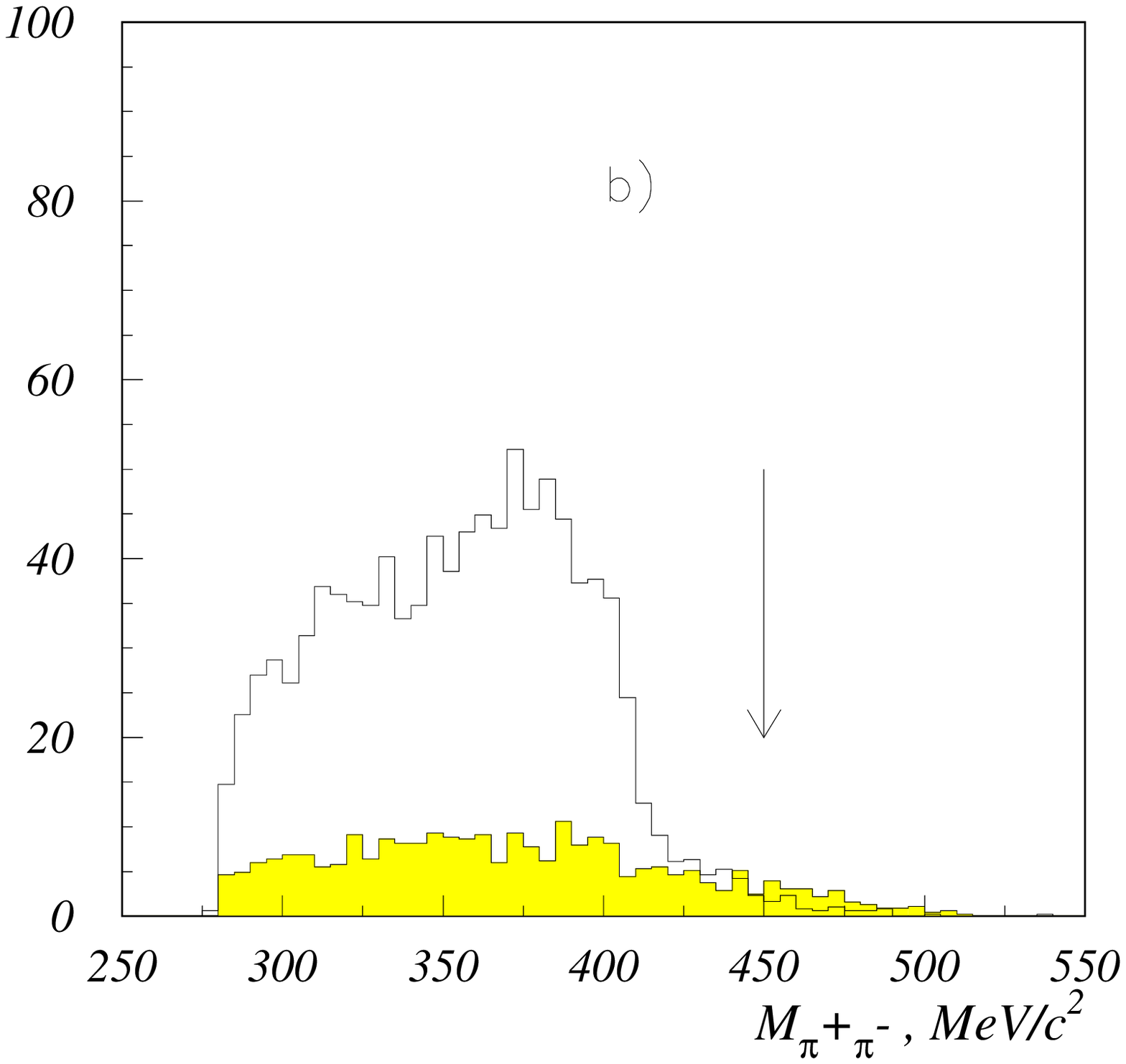}
  \\
  \caption{The invariant mass $M_{\pi^+\pi^-}$. a)
  Points with error bars --- data; transparent histogram --- MC of
    the process (\ref{eq:kskl}); hatched histogram --- MC of the
    reaction (\ref{eq:kpkm}); b) Simulation of the $\phi \to \etapg$
    decay; transparent histogram --- decay $\eta \to
    \pi^+\pi^-\pi^0$; hatched histogram --- decay $\eta \to
    \pi^+\pi^-\gamma$} \label{fig:etapg-ppim}
\end{center}
\end{figure}

About 250 events were selected after application of the criteria
described above. For these events the distribution of the invariant
mass of the pairs of charged pions $M_{\pi^+\pi^-}$ is shown in
Fig.\ref{fig:etapg-ppim}a) by the black squares with error bars
(four entries from one event). 
The transparent histogram in this Figure presents the distribution
obtained by the complete Monte Carlo simulation (MC) of the CMD-2 
detector \cite{cmd2sim} for
the process (\ref{eq:kskl}), while the hatched histogram is the MC of
the reaction (\ref{eq:kpkm}). The number of events in \ks\kl\ MC
samples corresponds to approximately the same integrated luminosity as
in our data. For comparison, in Fig.\ref{fig:etapg-ppim}b)
the same distributions of $M_{\pi^+\pi^-}$ are shown for the
simulation of the decays (\ref{eq:3pi}) and (\ref{eq:ppg}).

As one can see from Fig. \ref{fig:etapg-ppim}a), the main
contribution to the selected events comes from the process $\phi \to
\ks\kl$ followed by the decays $\ks \to \pi^+\pi^-$ and $\kl \to
\pi^+\pi^-\pi^0$. The requirement that  $M_{\pi^+\pi^-} < 450
\,\mbox{MeV}/c^2$ for each of the four possible $\pi^+\pi^-$
pairs efficiently rejects the \ks\kl\ events and
does not influence the events of the decay under study 
(see Fig. \ref{fig:etapg-ppim}b)). After that
only 23 events survive.

\begin{figure}[ht]
\begin{center}
  \includegraphics[width=0.49\textwidth]{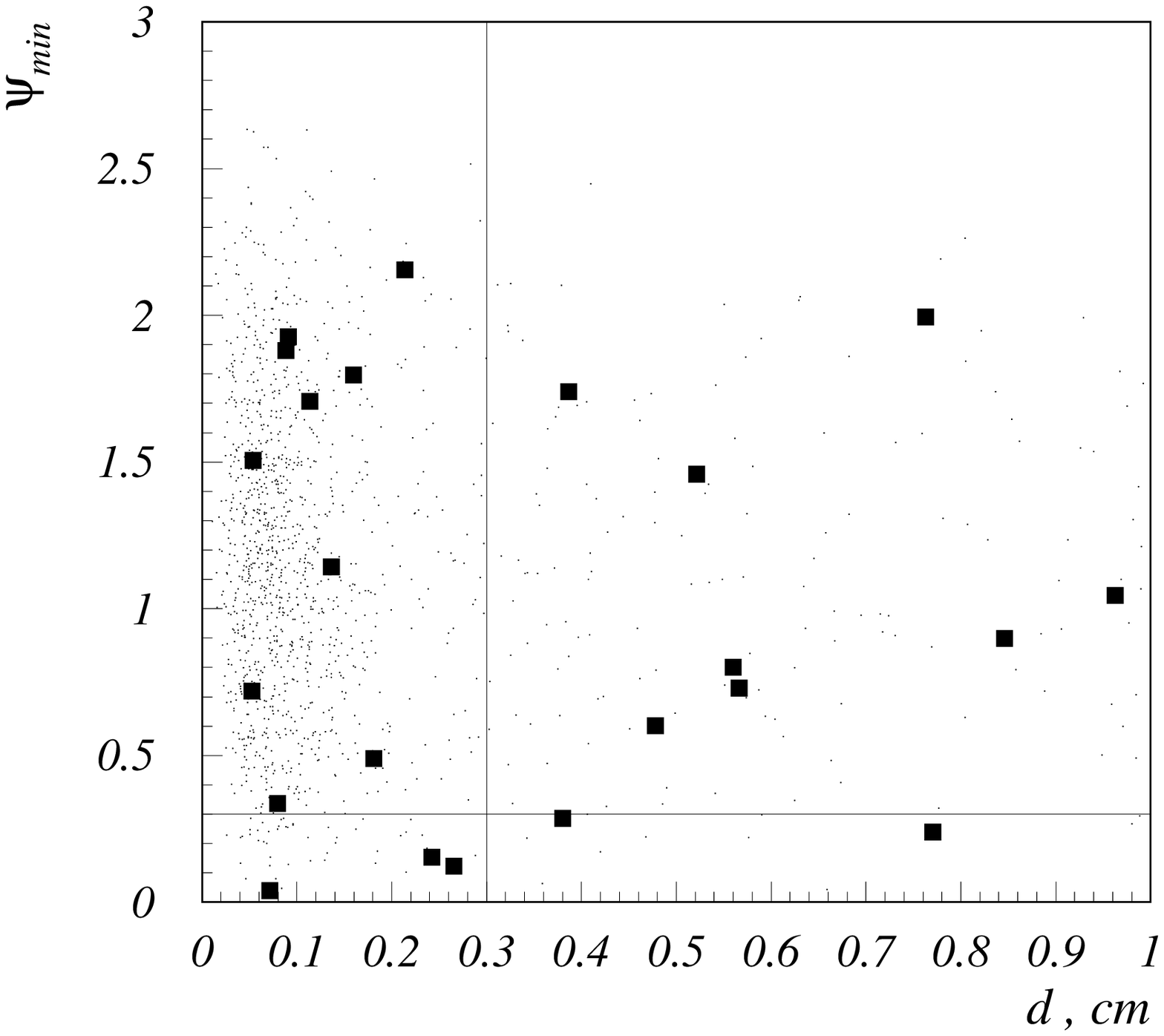}
  \hfill
  \includegraphics[width=0.49\textwidth]{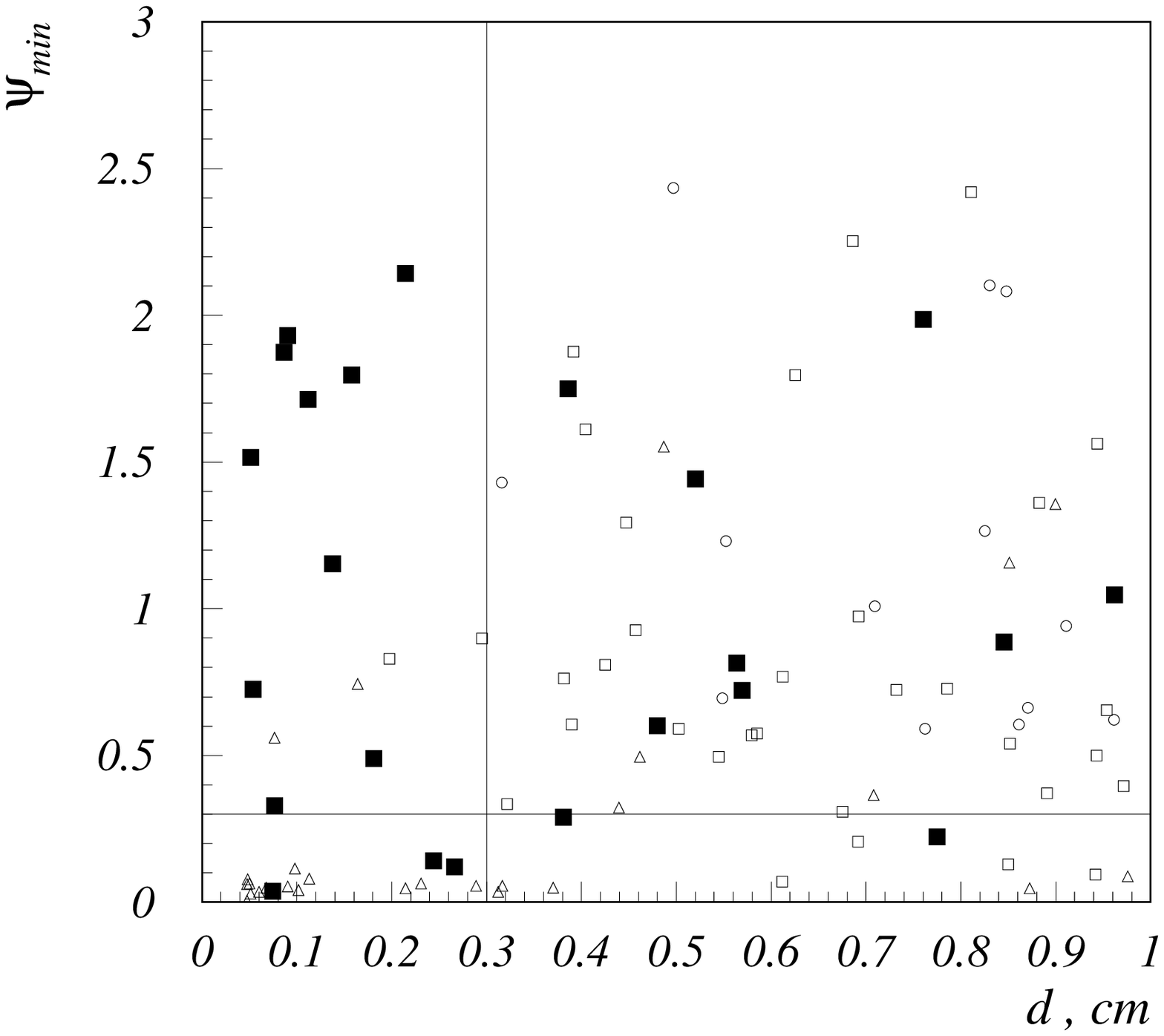}
  \\
  \caption{The minimum space angle between the tracks versus the 
  maximum impact parameter of the tracks. Black
  squares in both Figures present the experimental data; a)
  points are the MC of the decays (\ref{eq:3pi}) and (\ref{eq:ppg});
  b) $\circ$ is the MC of \ks\kl\ events; $\square$ is the
  MC of $K^+K^-$ events; $\triangle$ is the MC of 
  $\omega\pi^0$ events. The number of events in the  MC samples of
  charged and neutral kaons 
  and the number of simulated $\omega\pi^0$ events exceed the
  number of events of these reactions expected at our
  integrated luminosity by a factor of 10 and 4 respectively}
  \label{fig:etapg-dpsi}
\end{center}
\end{figure}

To complete separation of the \etapg\ events and estimate the
remaining background from the $\omega\pi^0$, \ks\kl\ and $K^+K^-$
events, the distributions of the minimum space angle between the
tracks and the maximum impact parameter of the tracks were analyzed
for 23 selected events. 
In Fig.\ref{fig:etapg-dpsi}a) the scatter plot of $\psi_{min}$ versus
$d$ is shown for the experimental data in comparison to the
simulation of the decays (\ref{eq:3pi}) and (\ref{eq:ppg}). In
Fig.\ref{fig:etapg-dpsi}b) the same experimental data are compared to
the MC of the background events \ks\kl\,, $K^+K^-$ and $\omega\pi^0$.

It is obvious from these Figures, that  events of the decay under
study contribute to the region of small impact parameters
and feature a broad distribution of $\psi_{min}$.
The background events $\omega\pi^0$ fall mainly in the region of small
angles: $\psi_{min} \lesssim 0.2$, while the events with neutral and
charged kaons have approximately flat distribution of the maximum
impact parameters of tracks with $d \gtrsim 0.3$ cm.
Correspondingly, the plane of the
parameters $\psi_{min}$ and $d$ was subdivided into three following regions:
\begin{enumerate}
\item the \etapg\ region: $\psi_{min} > 0.3$ and $d < 0.3$ cm;
\item the $\omega\pi^0$ region: $\psi_{min} < 0.3$ and $d < 0.3$ cm;
\item the $K\bar{K}$ region: $\psi_{min} > 0.3$ and $d > 0.3$ cm.
\end{enumerate}
The fourth region ($\psi_{min} > 0.3$ and $d > 0.3$ cm) contains 
a small contribution from all three background reactions.

\begin{figure}
\begin{center}
  \includegraphics[width=0.45\textwidth]{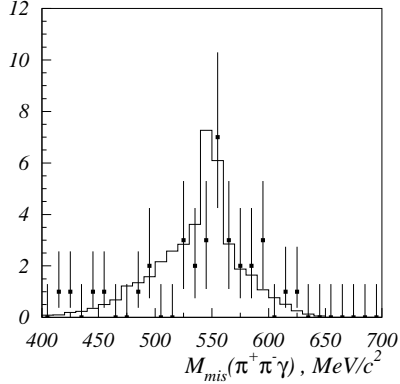}
  \caption{The missing mass  for the pair of pions
    and the photon with the energy closest to that of the
    monochromatic photon from the decay $\phi \to \etapg$;
    histogram --- Monte Carlo, black squares with error bars
    --- data}
  \label{fig:etap-2pgrm}
\end{center}
\end{figure}

Ten candidate events of the $\phi \to \etap$ decay were selected in
the \etapg\ region: $\tilde{N}_{\etapg} = 10$.
For these events the distribution of the missing mass for the pair of
charged pions and the photon with the energy closest to that of the
monochromatic photon from the decay $\phi \to \etapg$ is shown in
Fig.~\ref{fig:etap-2pgrm}. The MC spectrum in this Figure was obtained
as a sum of the contributions of the processes (\ref{eq:3pi}) and
(\ref{eq:ppg}) with the numbers of events proportional to the
corresponding decay probabilities. One can see that the experimental
data agree well with the simulation.

The number of background events was found to be:
$\tilde{N}_{\omega\pi^0} = 3$ and $\tilde{N}_{K\bar{K}} = 8$ in the
regions $\omega\pi^0$ and $K\bar{K}$ respectively.

\begin{figure}
  \begin{center}
    \includegraphics[width=\textwidth]{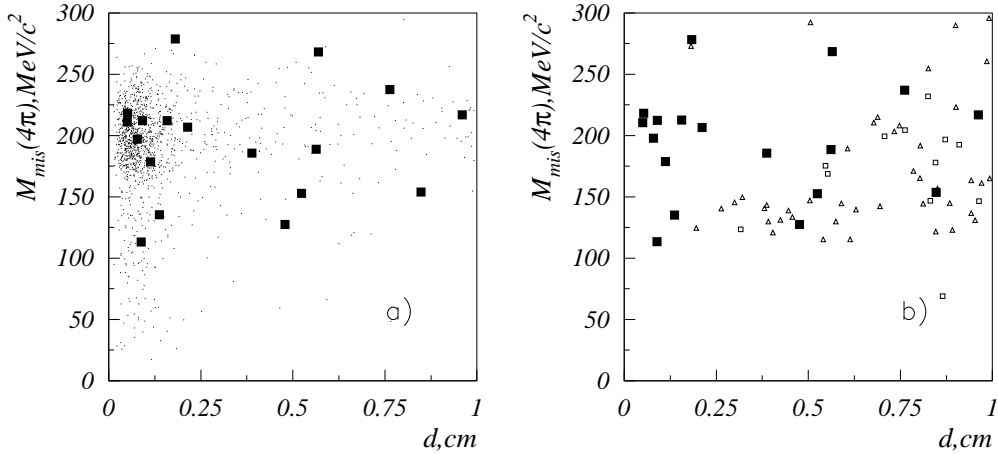}
    \caption{The missing mass of $4\pi$ versus the
      maximum impact parameter of the tracks. The black squares in 
      both pictures -- experimental data; (a) dots -- sum of Monte
      Carlo of the processes (\ref{eq:3pi}) and (\ref{eq:ppg}); (b)
      light squares -- MC of the $\phi \to \ks \kl$ decay, light
      triangles -- MC of the $\phi \to K^+ K^-$ decay} 
    \label{fig:etap-4pirm}
  \end{center}
\end{figure}

Figure \ref{fig:etap-4pirm}a) shows the distribution of the missing
mass of all four charged pions versus the maximum impact
parameter of the tracks for the events, which were selected under the
condition $\psi_{min} > 0.3$. Good agreement is observed  between
the Monte Carlo simulation of the signal and the experimental data for
$d < 0.3$ cm.
In Fig.~\ref{fig:etap-4pirm}b the same distribution of the
experimental data is shown in comparison to the simulation of the
background processes (\ref{eq:kskl}) and (\ref{eq:kpkm}).
These Figures provide one more verification of our assumption that the
\etapg\ events contribute to the region of small impact parameters,
while events with the decays of neutral and charged kaons have 
tracks which originate at rather large distances from the
interaction point.

The following probabilities to observe the background events
$\omega\pi^0$, \ks\kl\ and $K^+K^-$ in the \etapg\ region,
were obtained from the MC:
\begin{displaymath}
W_{\omega\pi^0} = \frac{N_{\omega\pi^0}(\psi_{min} > 0.3\,, d <
  0.3\,\mbox{cm})}{N_{\omega\pi^0}(\psi_{min} < 0.3\,, d <
  0.3\,\mbox{cm})} = 0.14 \pm 0.10\,,
\end{displaymath}
\begin{displaymath}
W_{K\bar{K}} = \frac{N_{K\bar{K}}(\psi_{min} > 0.3\,, d <
  0.3\,\mbox{cm})}{N_{K\bar{K}}(\psi_{min} > 0.3\,, d >
  0.3\,\mbox{cm})} = 0.10 \pm 0.05\,.
\end{displaymath}
Using these probabilities and the numbers of events
$\tilde{N}_{\omega\pi^0}$ and $\tilde{N}_{K\bar{K}}$, the background
contribution into ten candidate events $\tilde{N}_{\etapg} $ was
estimated. Finally, the number of the \etapg\ events was calculated
from the formula:
\begin{displaymath}
N_{\etapg} = \tilde{N}_{\etapg} - \tilde{N}_{K\bar{K}} W_{K\bar{K}} -
\tilde{N}_{\omega\pi^0} W_{\omega\pi^0} = 8.8 ^{+3.8}_{-3.2}\,.
\end{displaymath}
The error is determined  by the statistics of selected events only, 
as well as the  background contribution 
whereas the errors of the probabilities W$_{\omega\pi^0}$ and W$_{K\bar{K}}$ 
will be included in the systematic uncertainty. 
Here we did not take into account the possibility to observe 
\etapg\ events in the background regions as well as the probabilities to
detect the $\omega\pi^0$ events in the region $K\bar{K}$ and vice versa.
All these effects contribute less that 3\% into $N_{\etapg}$ and were
also included into the systematic uncertainty.

The detection efficiency was obtained using the 
Monte Carlo simulation of  the
processes (\ref{eq:3pi}) and (\ref{eq:ppg}): $\epsilon_{3\pi} =
0.097 \pm 0.003$ and $\epsilon_{\pi\pi\gamma}= 0.091 \pm 0.003$.

The process (\ref{eq:kskl}) 
was also used for the normalization and calculation of
the $\phi \to \etapg$ decay probability.
To select $\ks \kl$ events, conditions similar to those for
\etapg\ were used with one exception. We required $M_{\pi^+\pi^-} >
450\,\mbox{MeV}/c^2$ for at least one $\pi^+\pi^-$ pair.
Note that this condition effectively suppresses the process
(\ref{eq:kpkm}). 
The selection efficiency was determined from MC: $\epsilon_{\ks
  \kl} =  ( 4.9 \pm 0.3 ) \cdot 10^{-4}$.
The number of detected events is $ N_{\ks \kl} = 216$.

The relative probability of the decay $ \phi \to \etapg $ was calculated
using the formula:
\begin{eqnarray}
&& \frac{Br(\phi \to \etapg)} {Br(\phi \to \ks\kl)} =
\frac{N_{\etapg}}{N_{\ks\kl}} \cdot
\frac{Br(\ks\to \pi^+\pi^-) Br(\kl \to \pi^+\pi^-\pi^0)}
{Br(\etap \to   \pi^+\pi^-\eta)} \nonumber \\
& \times & \frac{\epsilon_{\ks\kl}}
{\epsilon_{3\pi} Br(\eta \to \pi^+\pi^-\pi^0) + 
 \epsilon_{\pi\pi\gamma} Br(\eta \to \pi^+\pi^-\gamma)}
\label{eq:br} \\
& = & (1.46 ^{+0.64}_{-0.54} \pm 0.18) \cdot 10^{-4} \;. \nonumber
\end{eqnarray}

The systematic uncertainty arises from the following contributions: 
6.7\% comes from the
uncertainties in determination of the detection efficiencies
$\epsilon_{3\pi}$, $\epsilon_{\pi\pi\gamma}$ and
$\epsilon_{\ks \kl}$; 7\% comes from the stability of the selection criteria,
6.4\% is due to the background subtraction
and 4.1\% comes from the uncertainties in the decay probabilities used in
the Eq.(\ref{eq:br}). All decay probabilities were taken from~\cite{pdg}.
The overall systematic uncertainty is 12.3\%.

Finally, taking the value of $Br( \phi \to \ks \kl)$ from~\cite{pdg}, the 
following branching ratio was obtained:
\begin{eqnarray} 
   Br(\phi \to \etapg) = (4.9 ^{+2.2}_{-1.8} \pm  0.6) \cdot
     10^{-5} \; .&& \nonumber
\end{eqnarray}

The energy behaviour of the cross section calculated using ten selected
events  does not contradict the shape expected for the sum of
the $\phi$ meson and non-resonant background.
 
\begin{table}
\caption{Summary of the $\phi \to \etapg$ decay rate
     measurements}\label{tab:sum} 
\begin{center}
\begin{tabular}{|@{}c@{}|@{}l@{}|@{}c@{}|@{}c@{}|@{}c@{}|}
\hline
Experiment & 
\begin{tabular}[t]{l}
\etapg\ decay \\
mode
\end{tabular} &
\begin{tabular}[t]{c}
$\frac{Br(\phi\to\etapg)}{Br(\phi\to\eta\gamma)}$, \\
$10^{-3}$
\end{tabular} &
\begin{tabular}[t]{c}
$\frac{Br(\phi\to\etapg)}{Br(\phi\to\ks\kl)}$, \\
$10^{-4}$
\end{tabular} &
\begin{tabular}[t]{c}
$Br(\phi\to\etapg)$, \\
$10^{-5}$
\end{tabular} \\
\hline
\hline
CMD-2 \cite{tanya1st} & 
\begin{tabular}[t]{l}
$\etap\to\pi^+\pi^-\eta$, \\
$\eta \to \gamma\gamma$
\end{tabular} & 
$9.5^{+5.2}_{-4.0} \pm 1.4$ & & $12.0 ^{+7.0}_{-5.0} \pm 1.8$ \\ 
\hline
SND \cite{snd} &
\begin{tabular}[t]{l}
$\etap\to\pi^+\pi^-\eta$, \\
$\eta\to\gamma\gamma$
\end{tabular} & & &
$6.7^{+3.4}_{-2.9} \pm 1.0$ \\
\hline
\begin{tabular}[t]{@{}l@{}}
CMD-2 \cite{tanya2nd}, \\
includes \cite{tanya1st}
\end{tabular} &
\begin{tabular}[t]{l}
$\etap\to\pi^+\pi^-\eta$, \\
$\eta\to\gamma\gamma$
\end{tabular} &
$6.5^{+1.7}_{-1.5}\pm 0.8$ & & $8.2^{+2.1}_{-1.9} \pm 1.1$ \\
\hline
KLOE \cite{kloe} &
\begin{tabular}[t]{l}
$\etap\to\pi^+\pi^-\eta$, \\
$\eta\to\gamma\gamma$
\end{tabular} &
$7.1\pm 1.6 \pm 0.3$ & & $8.9 \pm 2.0 \pm 0.6$ \\
\cline{2-5}
&
\begin{tabular}[t]{l}
$\etap\to\pi^0\pi^0\eta$, \\
$\eta\to\gamma\gamma$
\end{tabular} &
$6.8^{+3.2}_{-2.5} \pm 0.9$ & & \\
\hline \hline
\begin{tabular}[t]{@{}l@{}}
CMD-2, \\ this work
\end{tabular} &
\begin{tabular}[t]{l}
$\etap\to\pi^+\pi^-\eta$, \\
$\eta\to\pi^+\pi^-\pi^0$ or \\
$\eta\to\pi^+\pi^-\gamma$
\end{tabular} & &
$1.46 ^{+0.64}_{-0.54} \pm 0.18$ & $4.9 ^{+2.2}_{-1.8} \pm 0.6$ \\
\hline
\end{tabular}
\end{center}
\end{table}

\section{Discussion}

The currently available data on the measurements 
of the $\phi \to \etapg$ decay probability are
shown in Table \ref{tab:sum} in comparison to our result.
The value of the decay probability $Br( \phi \to \etapg )$ obtained in
our work is lower than that in \cite{snd,tanya2nd,kloe}, but
is consistent with their results within a statistical uncertainty.

The previous CMD-2 measurement of $Br(\phi\to\etapg)$ in the decay chain 
$\phi \to \etapg$, $\etap \to \pi^+\pi^-\eta$ and $\eta \to \gamma\gamma$
\cite{tanya2nd} is based on the data set, which is statistically
independent of our data sample. Analysis of the sources of systematic
errors in both measurements also shows that they are uncorrelated. Thus,
to improve the accuracy of the branching ratio determination,  
we can combine both results:
\begin{displaymath}
Br(\phi\to\etapg)_{\mbox{CMD-2}} = (6.4 \pm 1.6) \cdot 10^{-5}\,.
\end{displaymath}

The authors are grateful to the staff of VEPP-2M for excellent
performance of the collider, to all engineers and technicians who
participated in the design, commissioning and operation of CMD-2.

\end{document}